\pdfoutput=1

\documentclass[11pt]{article}
\newcommand{\bench}{\textsc{TransRepo-bench}~}

\newcommand{\approach}{Skeleton-Guided-Translation}

\newcommand*{\secref}[1]{\S\ref{sec: #1}}
\newcommand*{\figref}[1]{Fig.~\ref{fig: #1}}

\usepackage{acl}

\usepackage{times}
\usepackage{latexsym}

\usepackage{booktabs}
\usepackage{colortbl}
\usepackage{xcolor}

\usepackage{listings}

\usepackage[T1]{fontenc}

\usepackage[utf8]{inputenc}

\usepackage{microtype}

\usepackage{inconsolata}

\usepackage{graphicx}
\usepackage{enumitem}
\title{\approach : A Benchmarking Framework for \\ Code Repository Translation with Fine-Grained Quality Evaluation}

\author{
  Xing Zhang\thanks{These authors contributed equally to this work.}$^2$\quad Jiaheng Wen\footnotemark[1]$^3$\quad Fangkai Yang$^1$\quad Pu Zhao$^1$\quad Yu Kang$^1$ \quad Junhao Wang$^4$\\ \textbf{Maoquan Wang$^1$\quad Yufan Huang$^1$\quad Elsie Nallipogu$^1$\quad Qingwei Lin$^1$,}\\ \textbf{Yingnong Dang$^1$\quad Saravan Rajmohan$^1$\quad Dongmei Zhang$^1$\quad Qi Zhang$^1$} \\
  $^1$Microsoft,
  $^2$Peking University,
  $^3$Zhejiang University,
  $^4$Tongji University\\
}

\begin{document}
\maketitle
\begin{abstract}
The advancement of large language models has intensified the need to modernize enterprise applications and migrate legacy systems to secure, versatile languages. However, existing code translation benchmarks primarily focus on individual functions, overlooking the complexities involved in translating entire repositories, such as maintaining inter-module coherence and managing dependencies. While some recent repository-level translation benchmarks attempt to address these challenges, they still face limitations, including poor maintainability and overly coarse evaluation granularity, which make them less developer-friendly.  

We introduce \approach, a framework for repository-level Java to C\# code translation with fine-grained quality evaluation. It uses a two-step process: first translating the repository’s structural ``skeletons,'' then translating the full repository guided by these skeletons. Building on this, we present \bench, a benchmark of high-quality open-source Java repositories and their corresponding C\# skeletons, including matching unit tests and build configurations. Our unit tests are fixed and can be applied across multiple or incremental translations without manual adjustments, enhancing automation and scalability in evaluations. Additionally, we develop fine-grained evaluation metrics that assess translation quality at the individual test case level, addressing traditional binary metrics' inability to distinguish when build failures cause all tests to fail. Evaluations using \bench\, highlight key challenges and advance more accurate repository-level code translation.
\end{abstract}
\section{Introduction}

Large language models are transforming software development, driving the need for enterprises to modernize systems and migrate legacy code to cloud-friendly languages. For example, migrating from C to Rust offers enhanced safety benefits \cite{Matsakis14}, and libraries like TensorFlow require synchronized updates across languages. However, existing code translation benchmarks fall short in addressing real-world complexities. Most focus on function-level tasks or competition-style problems \cite{yan23, lu21, khan23}, which, while foundational, fail to capture the challenges of translating entire repositories. Repository-level translation is critical for managing interconnected components, dependencies, and structural integrity \cite{jiao23}, making reliable benchmarks essential to evaluate model performance in these scenarios.


\begin{figure}[t]
  \centering
\includegraphics[width=0.48\textwidth]{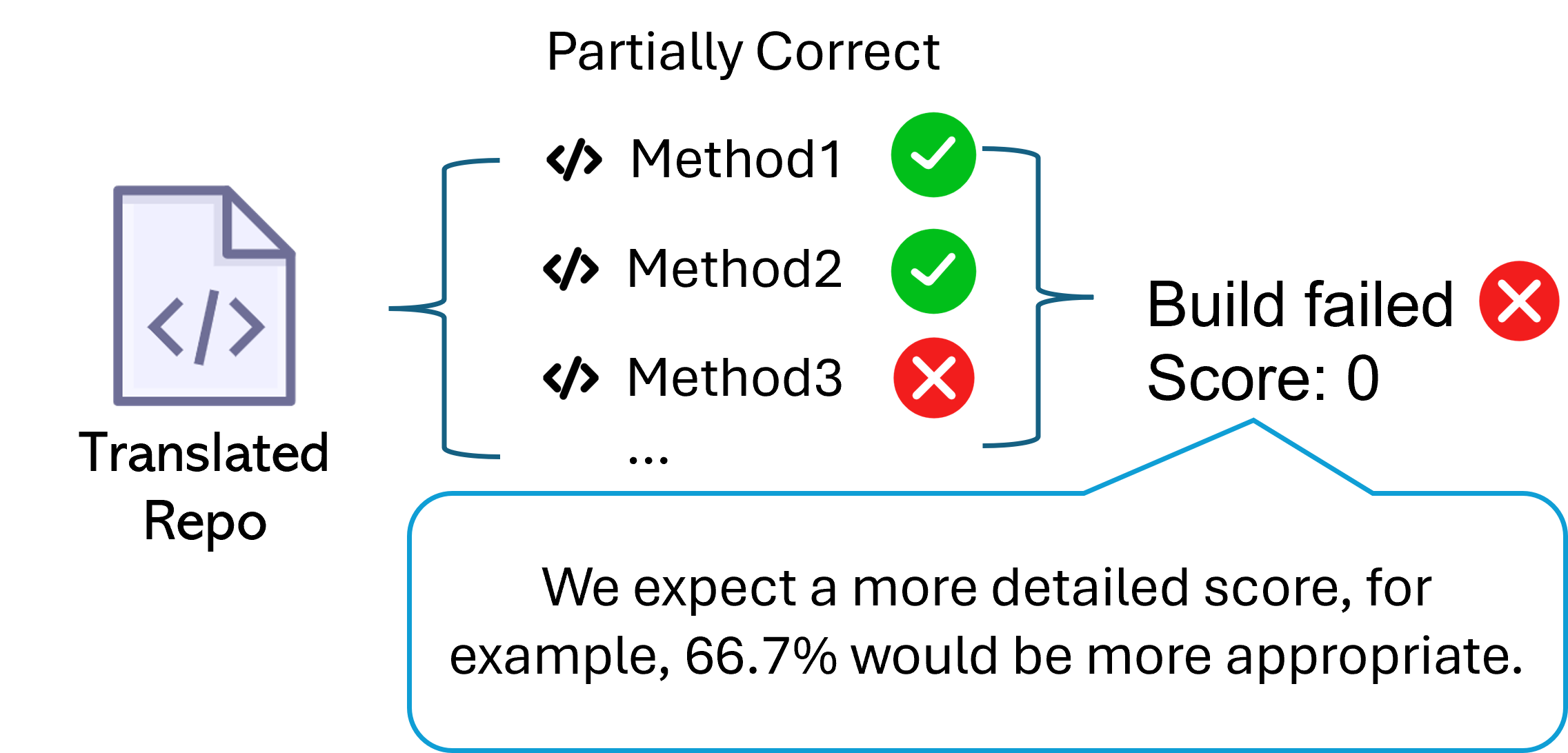}
  \caption{A more detailed quality evaluation to evaluate translated repositories is needed.}
  \label{fig: motivation}
\end{figure}

A key challenge in building a repository-level code translation benchmark is the absence of a systematic framework. For example, when updating part of a large Java-based SDK with new features, developers often cannot incrementally apply those changes to a corresponding C++ codebase without re-translating major portions. This lack of fine-grained control hinders maintainability, as small updates become disproportionately costly. A robust framework is required to accommodate partial updates and reduce overhead, ensuring that evolving codebases can be maintained efficiently across multiple languages without constant full-scale retranslation.

Another significant challenge is the scarcity of library-level parallel corpora, which makes ensuring correctness through automated testing difficult. Literal line-by-line verification methods, such as codeBLEU \cite{ren20}, do not guarantee functional validation of the translated code. Additionally, automatic generation of functional tests for translated projects is still immature and unreliable \cite{eniser24}. A more feasible approach is to translate the unit tests from the source library into the target language, enabling systematic validation of the translated library’s functionality. However, this method introduces concerns about the accuracy of the translated tests and the necessity to maintain consistency with the translated source code’s interfaces. Ensuring both the correctness of the translated tests and their alignment with the translated code interfaces is crucial for reliable functional verification.

Finally, current evaluation metrics often fail to capture nuanced translation outcomes, leading to low usability for developers. RepoTransBench \cite{wang24}, for instance, focuses on a binary build success metric, offering limited insight into partial successes.
As illustrated in Figure \ref{fig: motivation}, this approach oversimplifies performance by ignoring partial successes, such as when some components of a repository translate correctly while others fail. Schaeffer et al. \cite{schaeffer23} caution against such nonlinear or threshold-based metrics, noting that they can create the illusion of sudden performance leaps. In contrast, continuous metrics, such as reporting the percentage of successfully translated modules (e.g., 66.7\%), not only enhance developer friendliness by clearly identifying where translations fail and guiding targeted fixes but also provide smoother and more predictable insights into model performance, thereby offering a better basis for assessing and forecasting model capabilities.

\subsubsection*{Our Contributions} 
Through overcoming these challenges, we present a novel framework, \approach, for translating code repositories with a focus on fine-grained quality evaluation. Our approach introduces a two-step process: first, we translate the repository skeleton to establish a clear structure and provide interfaces for further translation; then, we populate the skeleton while indexing dependencies for unit tests. This systematic method ensures consistency, facilitates targeted evaluation, and addresses the challenges of validating translations at both the structural and functional levels. Building on this approach, we introduce \bench, a novel benchmark tailored for repository-level code translation, which overcomes the limitations of existing benchmarks by leveraging our skeleton-based translation framework to enable precise evaluation.
Specifically:
\begin{itemize}[leftmargin=*]
    \item Framework for Both Repository Translation and Fine-Grained Quality Evaluation: We present an a novel translation framework, \approach, with fine-grained evaluation metrics, for code repository translation. \approach\, employs a two-step process to extract and translate repository skeletons, capturing the structural essence and guiding the full translation to ensure consistency and address repository-level challenges such as cross-file dependencies and module interactions. Complementing this framework, our benchmark \bench provides detailed quality evaluation by scoring individual test cases based on unit tests and their associated code, offering more meaningful feedback than binary metrics like compilation success.


    \item High-Quality Open-Source Repository Benchmark: \bench~ includes high-quality open-source Java libraries and their translated counterparts in C\#, complete with unit tests and test configuration files for evaluation. The benchmark is constructed for both translation and fine-grained quality evaluation of code repositories, enabling researchers to evaluate models in practical scenarios that reflect repository-level demands.

    \item Evaluation of State-of-the-Art Models: \bench~ is validated through extensive evaluations of classic and state-of-the-art models and agents implemented with those models, providing a detailed analysis of their performance. This highlights key challenges in repository-level translation and identifies strengths and weaknesses of current models.
\end{itemize}

\begin{figure*}[ht]
  \centering
  \includegraphics[width=0.95\textwidth]{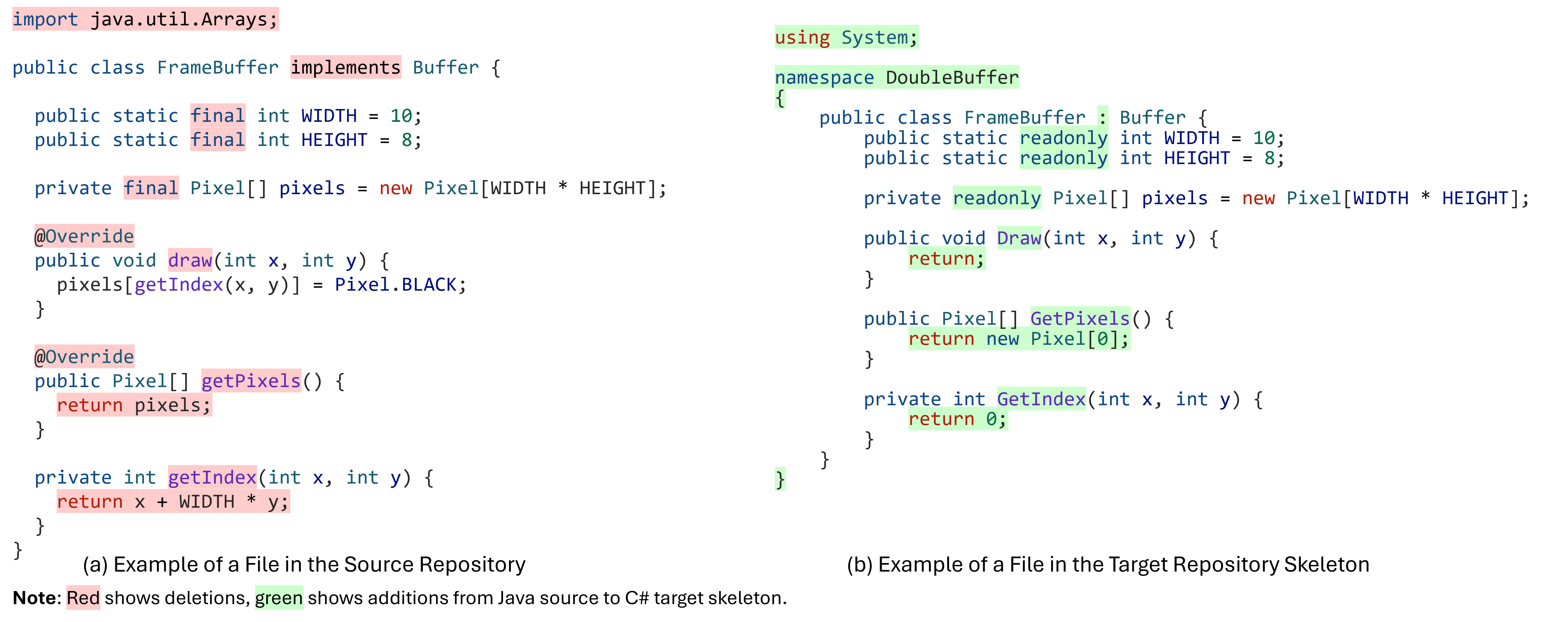}
  \caption{Input of Translation Task.}
  \label{fig: trans-input}
\end{figure*}

\section{Motivating Example}
In this section, we use an example to illustrate the challenges involved in building a repository-level code translation benchmark and explain our solutions more effectively.

\subsection{Challenges in Repository Translation and Quality Evaluation}

\emph{Lack of a Systematic Translation Framework.} 
The partial Java-to-C\# translation shown in Figure \ref{fig: trans-input} highlights the challenges posed by incremental updates when using LLMs for translation and underscores the necessity of a systematic framework. In the Java snippet in Figure \ref{fig: trans-input}(a), the FrameBuffer class correctly calculates indices and renders pixels. However, if a developer adds a new method to the FrameBuffer class, the absence of a structured translation framework would require re-translating significant portions of the code whenever the original Java module changes—drastically reducing maintainability. 

\emph{Lack of Parallel Corpora.}
Repository-level translations face challenges due to misaligned source and target files, making it hard to verify correctness across languages. For example, translating Java code (Figure \ref{fig: trans-input}(a)) to C\# is difficult without corresponding C\# tests, as there’s no direct way to compare behaviors or outcomes, especially for complex logic or edge cases. One solution is translating high-coverage Java tests into equivalent C\# tests, but this raises another issue: how to efficiently verify that the translated tests maintain the original intent, coverage, and reliability. Without this, testing inconsistencies could undermine confidence in the translation.

\emph{Lack of a Fine-Grained Evaluation Metric.}
An overreliance on coarse metrics (e.g., whether a repository ``builds'' at all) limits developers' ability to identify issues in translated code. For example, if the Draw method is translated incorrectly by calling getIndex instead of GetIndex, the code fails to compile, making it impossible to evaluate the translation quality of other functions. However, methods like GetPixels and GetIndex might have been translated correctly. This binary pass/fail approach obscures partial successes and forces developers to manually search for issues. More granular metrics—such as module-level correctness or individual function fidelity—would help developers locate problematic sections more efficiently, streamlining debugging and refinement.

\begin{figure*}[ht]
  \centering
  \includegraphics[width=\textwidth]{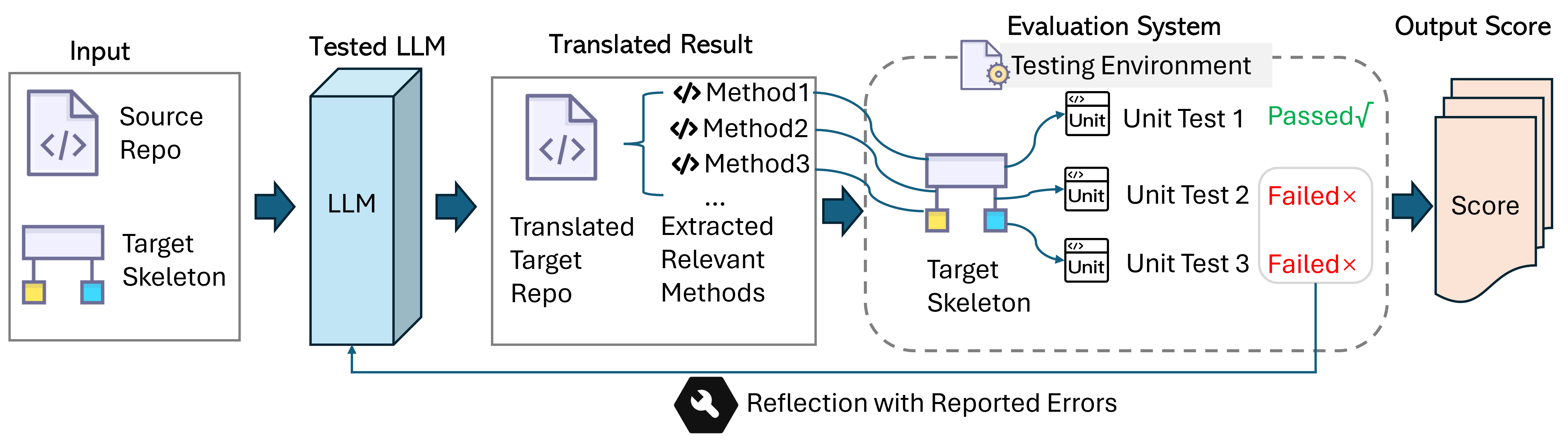}
  \caption{Framework of Our Evaluator.}
  \label{fig: eval-frame}
\end{figure*}

\subsection{Solution: Standardizing Code Repository Translation with Partial Evaluation}
Figure \ref{fig: eval-frame} illustrates our solution overcoming the above challenges. To ensure consistency between the translation process and the testing of the target repository, as well as to achieve fine-grained evaluation of individual unit tests, we propose introducing a ``skeleton'' of the target repository during the translation process. This skeleton serves to guide large language models (LLMs) in focusing on the accuracy of dependencies and interface translations within the code. Additionally, the skeleton can be incrementally filled with partial translation results, enabling fine-grained execution-based evaluation of translation quality. 

\emph{Facilitating Maintainability.} 
Figure \ref{fig: trans-input}(b) demonstrates our framework, Skeleton-Guided-Translation, where the C\# code serves as a ``target repository skeleton''. Unlike the fully translated Java code in Figure \ref{fig: trans-input}(a), this skeleton establishes consistent interfaces while leaving method bodies mostly empty or trivial. 
This systematic methodology underpins maintainability: whenever the Java repository evolves, only the corresponding skeleton sections in C\# require incremental updates, avoiding the need to overhaul the entire translation.

\emph{Enhancing Testability.} 
Building on these skeletons significantly improves testability. Because the structural and interface definitions in both repositories match, any unit tests originally designed for the Java code—especially those focusing on API behavior—can be adapted to validate the C\# skeleton. 
Even if a method’s implementation in C\# is just a placeholder, the test environment can still verify that calls are made correctly and interfaces remain consistent. 

\emph{Improving Usability.} 
The framework’s finer-grained control over each translated module enhances developer usability. For instance, if the Draw method is translated incorrectly and fails to compile, unit tests that only call GetPixels and GetIndex can still be executed by copying their contents into the skeleton (as shown in the evaluation system in Figure \ref{fig: eval-frame}). This approach allows developers to verify that GetPixels and GetIndex were translated correctly, even when there are compilation errors elsewhere. Unlike coarse build-or-fail metrics, which obscure partial successes and force teams to search the entire codebase for issues, skeleton-based testing enables comprehensive evaluation of all translated content.

\section{\bench~ Benchmark}
In our benchmark, the source repository and target repository skeleton are provided for users, aiming for LLMs to generate a complete target repository. The correctness of the generated repository is verified using the target repository's unit tests within the designated testing environment. In this section, we begin by presenting the benchmark content, followed by detailing the construction process of \bench, and conclude with an introduction to our fine-grained evaluation design.


\subsection{Benchmark Overview}
Each code translation task in \bench~ consists of a source repository and its evaluation setup, structured as <source repository, target repository skeleton, target repository unit tests, testing environment>. Currently, we use Java to C\# translation as an example, with plans to extend the benchmark to include additional language pairs in the future.

As illustrated in Figure \ref{fig: trans-input}, the input for the translation task comprises source repositories, written in Java, serving as the code bases to be translated, and a target repository skeleton, which acts as a crucial contract for effective evaluation, supporting both the translation and evaluation phases. 
It is a highly simplified version of the target repository, where all functions are replaced with trivial implementations (e.g., a single return statement). This skeleton preserves the original repository's file structure, dependencies, and static values while ensuring it compiles successfully.
The evaluation setup comprises the unit tests for the target repository and the necessary testing configuration files to execute those tests.

\begin{figure}[t]
  \centering
  \includegraphics[width=0.48\textwidth]{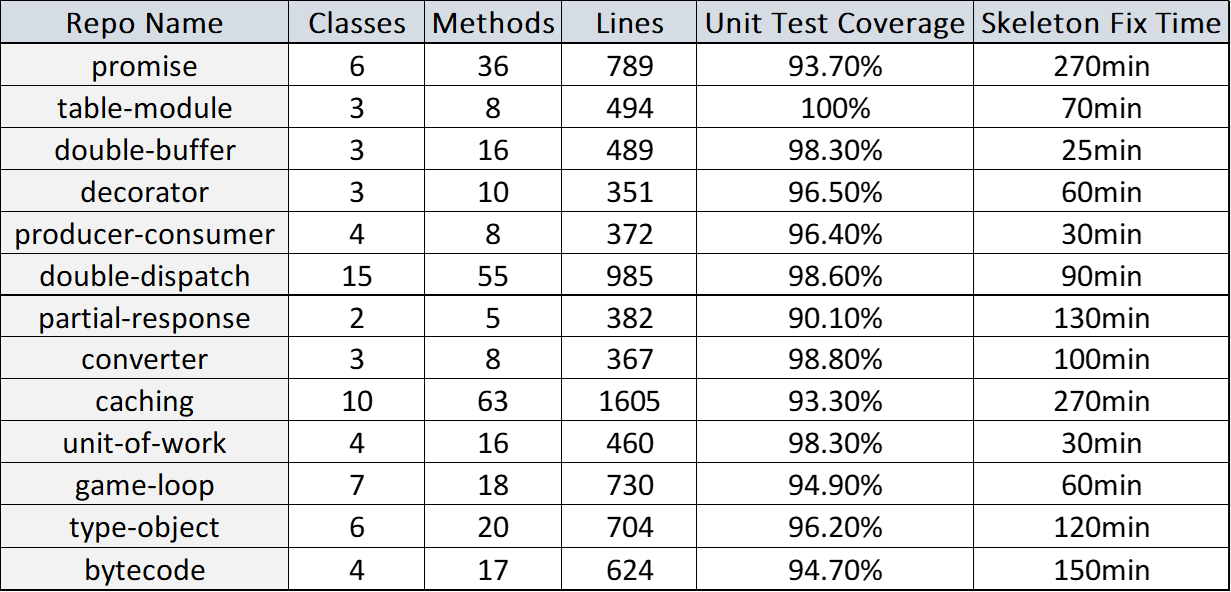}
  \caption{Resulting Benchmark}
  \label{fig: bench-summary}
\end{figure}

\bench~ comprises 13 tasks for translating code repositories. Detailed information is provided in Figure \ref{fig: bench-summary}. It outlines key features for each repository, such as the number of classes, methods, and lines of code, along with unit test coverage percentages. The data demonstrates a diverse range of repository complexities, from small repositories with minimal classes and methods to larger ones with hundreds of methods and significant code coverage. This diversity underscores the robustness and variety of tasks encompassed by \bench, ensuring comprehensive evaluation in repository translation tasks.
\begin{figure}[t]
  \centering
  \includegraphics[width=0.3\textwidth]{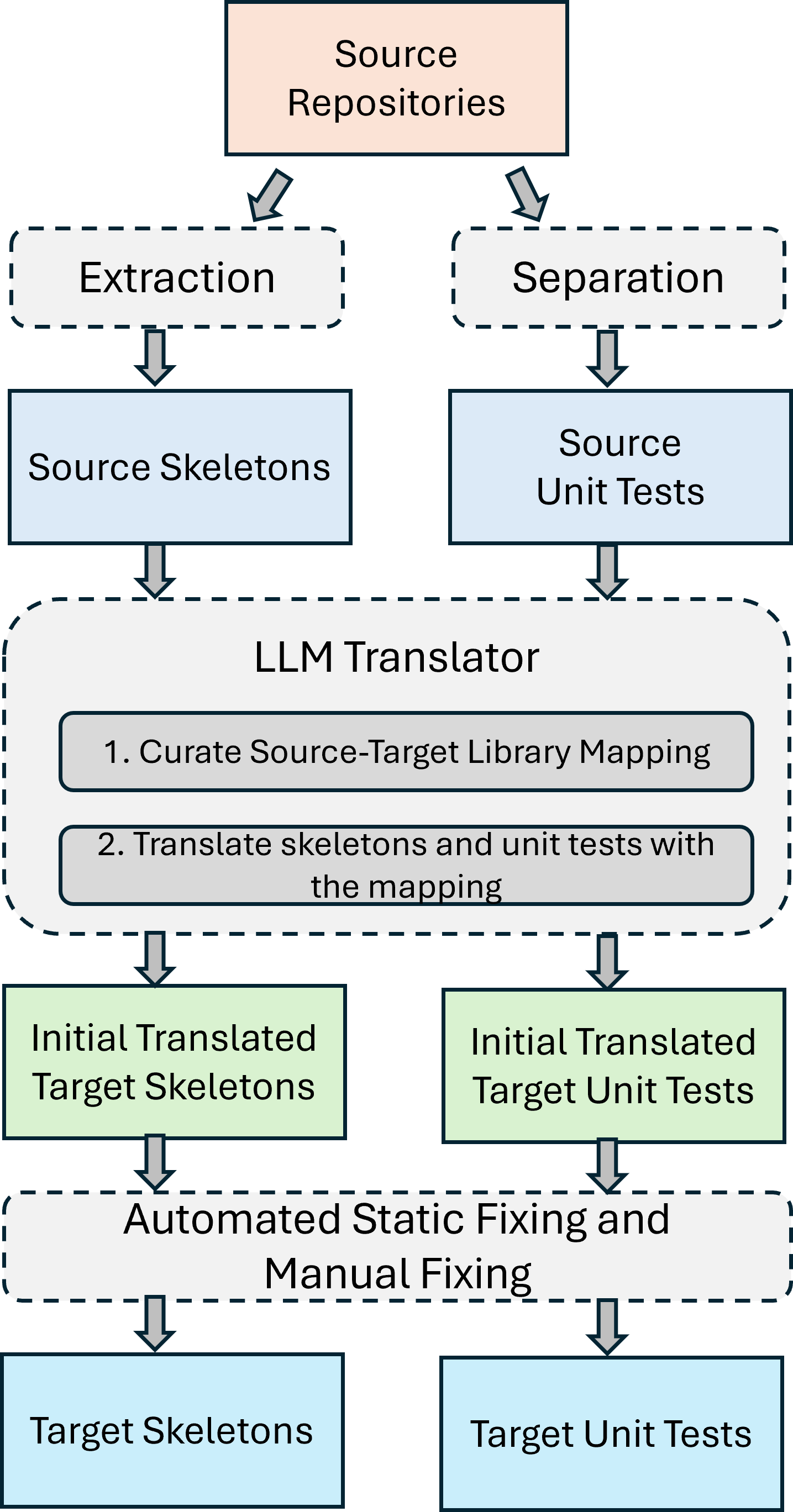}
  \caption{Framework of the Benchmark Construction.}
  \label{fig: bench-constr}
\end{figure}

\subsection{Benchmark Construction}
This section provides a detailed explanation of the benchmark construction process, as summarized in \figref{bench-constr}. We begin by describing the collection of source repository datasets (\secref{dataset}). Next, we outline the method for extracting the skeletons of source repositories and translating them into the target language (\secref{skeletons}). We then detail the process of obtaining unit tests for the target repositories (\secref{test_trans}). Finally, we describe the preparation of testing environments for each repository (\secref{test_env}).

\subsubsection{Source Repository Collection}\label{sec: dataset}
The dataset for the source repository is curated from open-source GitHub projects, following these selection criteria: (1) repositories must have over 100 stars; (2) they must include a testing workflow; and (3) their testing workflow must execute successfully and pass when run locally.

Given that previous attempts \cite{Pan24} to translate entire repositories often failed to even compile, we decided to start with common and mature repositories.
We selected ``java-design-patterns,'' a Java library that provides a comprehensive collection of design pattern implementations. This choice was made because libraries of this nature typically feature higher code quality, comprehensive testing, and successful test execution. 

\subsubsection{Skeleton Extraction and Translation}\label{sec: skeletons}
Repository skeletons are simplified versions of repositories where all function implementations in files (excluding testing files) written in the source language are replaced with trivial return statements. Files written in other languages are retained, ensuring that the skeleton can still compile and execute successfully. These skeletons preserve the original file structure, dependencies, interfaces, and static values of the source repository.

Specifically, function bodies are replaced with return statements that provide trivial values corresponding to the output types of the functions, enabling successful compilation by satisfying type-checking requirements. For example, if a function returns an int, we replace its body with ``\texttt{return 0;}''; for functions returning objects, we use ``\texttt{return null;}''. For class constructors that have no return type, the function body is left empty. Similarly, for static blocks, only the assignments are retained, while the rest of the block is removed.

After extracting the skeletons, we translate them into the target language using a large language model (GPT-4o). However, most translated skeletons fail to compile successfully, necessitating significant manual effort to correct translation errors and restore proper functionality. As illustrated in Figure \ref{fig: bench-summary}, the column labeled ``Skeleton Fix Time'' captures the time spent on fixing the translated target repository skeletons. This metric reflects the approximate manual effort required to produce a target repository skeleton, serving as the foundation step in the entire repository translation process.

\subsubsection{Unit Test Translation}\label{sec: test_trans}

We translate the unit tests within source repositories into the target language using a large language model (GPT-4o) and the NUnit testing framework. However, the translated unit tests often fail to compile successfully, necessitating significant manual effort to resolve translation errors and ensure that the tests correctly interact with the source code they are intended to validate.

\subsubsection{Testing Environment Construction}\label{sec: test_env}
Building a testing environment involves creating a setup where unit tests can be executed, typically by specifying a Docker image, installing the necessary dependencies, and running the tests. For our process, we prepare a build configuration file in YAML format for the translated C\# project, based on the existing build file from the original Java project.

This step is primarily done manually, referencing the translated C\# skeleton to configure the build file. Additionally, we leverage a large language model (such as GPT-4o) to assist in translating the Java build file directly into a corresponding C\# build file. The translated build file is then refined and corrected to ensure it functions as intended for the C\# project.

To reduce the manual effort and facilitate the broader use of our framework, we have released several supporting resources: static repair scripts for skeletons and unit tests and automated configuration scripts for C\# projects. These tools significantly lower the barrier to adoption and improve efficiency for researchers and developers. The capabilities of the automatic repair scripts are currently limited, which required us to invest significant manual effort during the process. 

\subsection{Fine-Grained Evaluation Metrics Design}
To provide a more fine-grained evaluation of user-translated code, we leverage unit tests to score the translated output. Previous attempts to translate entire repositories often ended in compilation failures, preventing even the execution of unit tests. Pan et al. \cite{Pan24} claim that 77.8\% of the failures in large-model translations are due to compilation errors. In their experiments on two translated repositories, tasks failed solely because of these errors, making it difficult to obtain a more detailed error analysis. This limitation potentially obscures translations that might be correct or valuable, hindering a comprehensive evaluation of the model's translation capabilities.

To mitigate the binary impact of ``compile success vs. failure'' on our evaluation, we extract the source code relevant to individual unit tests and execute them within a guaranteed-compilable skeleton. During evaluation, the translated functions related to a specific unit test are copied into this skeleton, and the dotnet build and dotnet test commands are executed. This method ensures fine-grained scoring, free from the broader impact of translation errors in unrelated parts of the repository. We provide dynamic instrumentation scripts for extracting relevant code from unit tests within the evaluation system.

Our evaluation employs two primary metrics: build success rate and unit test success rate. The build success rate is calculated as the proportion of unit tests that compile successfully out of all unit tests for each library. Similarly, the unit test success rate measures the proportion of unit tests that pass successfully out of those that compile. For an overall assessment of a model's performance, we compute the average of these scores across all evaluated libraries. This approach allows unit tests to execute successfully while isolating evaluation from compilation errors in unrelated parts of the translated repository. The key challenge lies in extracting the source code relevant to each unit test. To address this, we instrument the Java source code at the function level to identify the code invoked by each unit test during execution. Using the structural mapping between the source Java and target C\# repositories, which share the same file and class structure, we then locate the corresponding C\# source code required to execute the unit tests.
\begin{figure}[t]
  \centering
\includegraphics[width=0.48\textwidth]{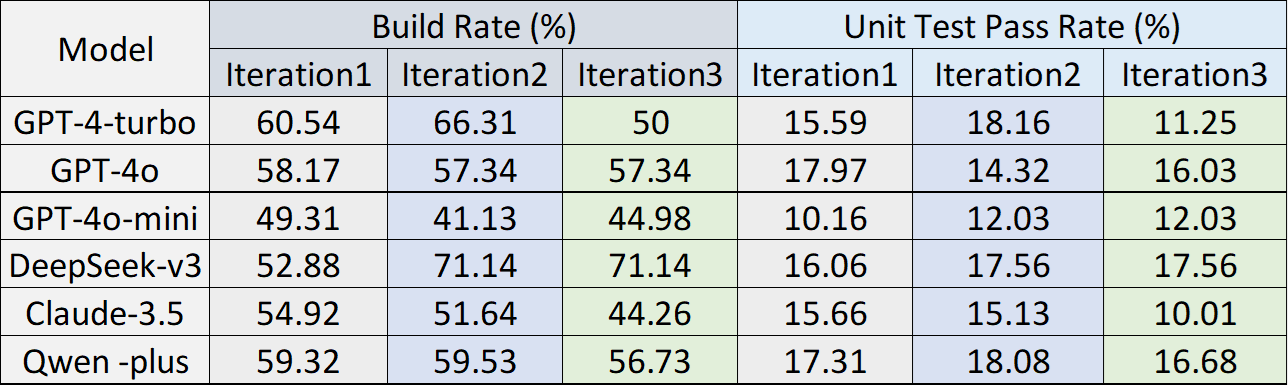}
  \caption{Build rates (\%) and Unit test pass rates (\%) for different repositories across various models.}
  \label{fig: overall-performance}
\end{figure}
\begin{figure*}[t]
  \centering
  \includegraphics[width=\textwidth]{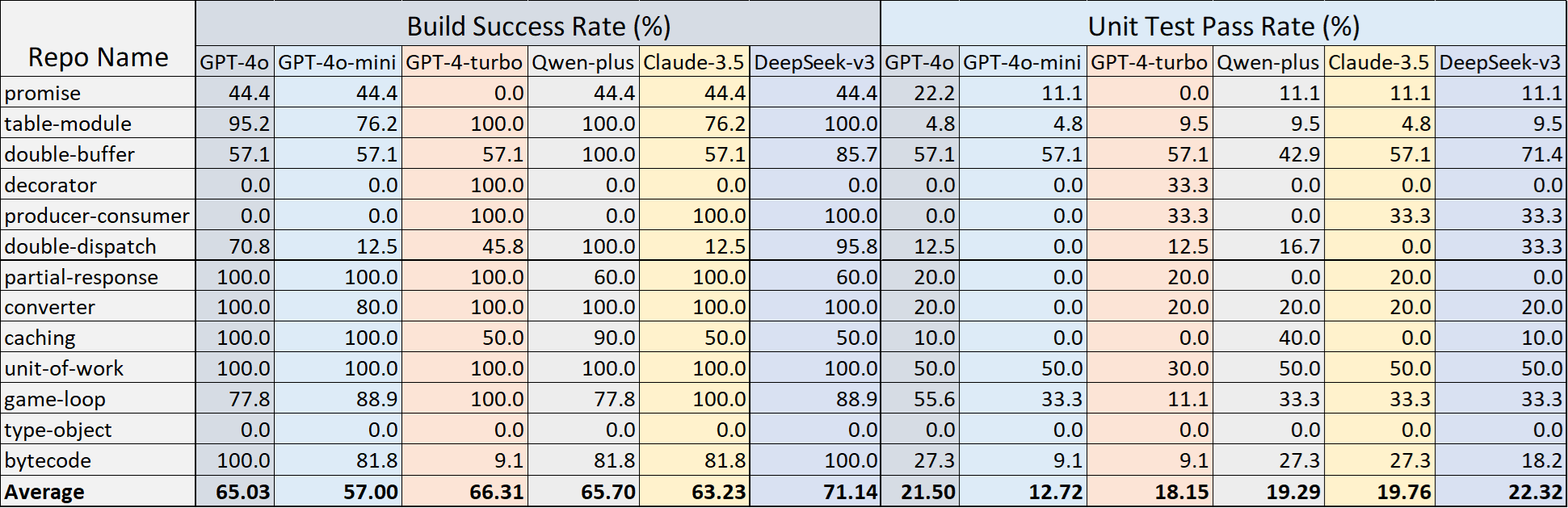}
  \caption{Build rates (\%) and Unit test pass rates (\%) for different repositories across various models.}
  \label{fig: build-test-rate}
\end{figure*}

\section{Evaluation}
We first analyze how LLMs perform on our benchmark. Then, we highlight the effectiveness of our novel framework, which incorporates target repository skeletons for translation and fine-grained evaluation.
\subsection{Model Performance on \bench}
We evaluate the performance of state-of-the-art LLMs on the task of translating code repositories from Java to C\#. Next, we conduct a failure analysis based on the experimental results. 
\subsubsection{Model Selection}
To evaluate the performance of state-of-the-art large language models (LLMs) on the code repository translation task, we selected six representative models: GPT-4o, GPT-4o-mini, GPT-4-turbo, Qwen-plus-1220, Claude-3.5-sonnet-20240620, and Deepseek-v3. GPT-4o, GPT-4o-mini, and GPT-4-turbo represent versatile general-purpose LLMs with strong capabilities in reasoning and language understanding, optimized for different levels of computational efficiency and application contexts. Qwen-plus-1220 and Claude-3.5-sonnet-20240620 are advanced models that bridge general-purpose tasks and specialized reasoning, offering nuanced language comprehension for complex scenarios. Deepseek-v3, on the other hand, is fine-tuned specifically for code-related tasks, focusing on programming language understanding and transformation.

\subsubsection{LLMs Performance}
Figure \ref{fig: overall-performance} presents the overall performance evaluation of various large language models (LLMs) across three iterations, focusing on two metrics: Build Rate and Unit Test Pass Rate. Notably, DeepSeek-v3 demonstrates a consistent improvement across iterations, achieving the highest Build Rate (71.14\%) and a competitive Unit Test Pass Rate (17.56\%) by the third iteration, showcasing its robustness and optimization. GPT-4-turbo starts strong with a Build Rate of 60.54\% in Iteration 1 but drops significantly to 50.00\% by Iteration 3, with its Unit Test Pass Rate also decreasing to 11.25\%. GPT-4o maintains a steady performance, with its Build Rate stabilizing at 57.34\% and a slight fluctuation in Unit Test Pass Rate, ending at 16.03\%. Models like GPT-4o-mini and Claude-3.5 exhibit weaker performance, with declining Build Rates and inconsistent trends in Unit Test Pass Rates. Overall, DeepSeek-v3 stands out as the most effective model, while others face challenges in sustaining performance.

The table results demonstrate that iterative refinement does not always lead to improved performance. One possible reason is the error propagation effect. In these scenarios, the outputs from previous iterations serve as the inputs for subsequent ones. If errors or inefficiencies are introduced in earlier stages, they may accumulate or intensify instead of being corrected. This is especially problematic if the models struggle to differentiate between constructive feedback and irrelevant noise during refinement.

\textbf{Build Rates.}
From the build success rates in Figure \ref{fig: build-test-rate}, DeepSeek-v3 exhibits the highest overall performance at 71.14\%, followed by GPT-4-turbo at 66.31\% and Qwen-plus at 65.70\%. Meanwhile, GPT-4o, GPT-4o-mini, and Claude-3.5 show slightly lower aggregate rates, at 65.03\%, 57.00\%, and 63.23\% respectively. Despite these average trends, there are notable variations across individual repositories. For instance, decorator and producer-consumer pose challenges for most models (with many yielding a 0\% build rate), whereas repositories such as converter, partial-response, and unit-of-work reach 100\% build success for multiple models. These discrepancies suggest that certain repository structures and coding patterns can significantly influence the success or failure of automatic translation.

\textbf{Unit Test Pass Rates. }
Figure \ref{fig: build-test-rate} highlights the unit test pass rates (\%) across various repositories for different large language models. 
In terms of unit test pass rates, DeepSeek-v3 again leads with an average of 22.32\%, slightly outperforming the next-best models: GPT-4o at 21.50\% and Claude-3.5 at 19.76\%. The remaining systems—GPT-4o-mini, GPT-4-turbo, and Qwen-plus—fall in a similar range, from 12.72\% to 19.29\%. Consistent with the build results, repository-level scores vary widely: while double-buffer and bytecode often pass many tests (notably exceeding 50\% in some cases), others such as producer-consumer and decorator register zero for nearly all models. The combination of low overall pass rates and stark repository-level contrasts highlights the complexity of fully preserving runtime behavior during translation, as even a successful build does not guarantee correct functionality across all test cases.

\subsubsection{Failure Analysis}
\begin{figure}[t]
  \centering
  \includegraphics[width=0.49\textwidth]{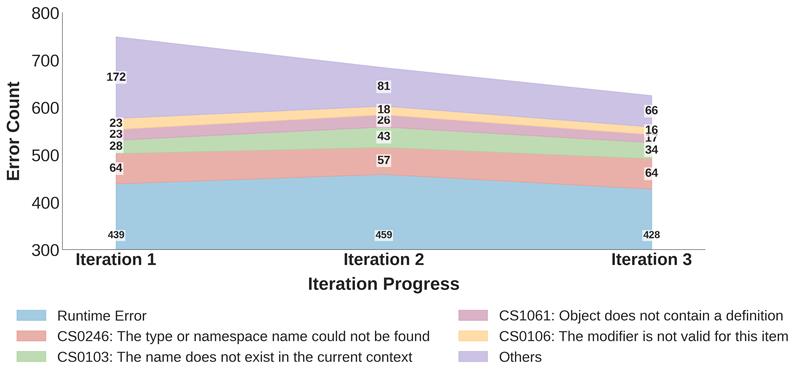}
  \caption{Changes in Error Proportions}
  \label{fig: error-analysis}
\end{figure}

\noindent\textbf{Overview.}
Figure~\ref{fig: error-analysis} illustrates the distribution and reduction of error types across three iterations, demonstrating the effectiveness of the iterative refinement process. The most frequent error category, Runtime Errors, decreased significantly from 439 in Iteration 1 to 428 in Iteration 3, highlighting its prominence and the consistent efforts to address it. Other common error types include \texttt{CS0246} (missing type or namespace), \texttt{CS1061} (missing member in an object), and \texttt{CS0103} (undefined variable or name), all of which exhibit similar trends of gradual reduction, indicating that the model effectively identifies and rectifies these errors over iterations. For instance, \texttt{CS0106} errors dropped from 23 to 16, while \texttt{CS1061} reduced from 23 to 17, respectively. The inconsistent decrease in \texttt{CS0103} and \texttt{CS0246} errors may result from the iterative process introducing new variables or dependencies without properly defining or importing them.

Additionally, less frequent error categories such as \texttt{CS0106} (modifier errors in method declarations) and ``Others'' demonstrate modest but steady reductions, signifying the model's capability to address a broader range of issues. The overall decline in total errors from 747 in Iteration 1 to 619 in Iteration 3 suggests that the model particularly excels at resolving syntactical and common logical errors, which tend to follow recognizable patterns. This result supports the claim that iterations are effective in error mitigation and highlights the strengths of large models in correcting repetitive, rule-based error types.

\noindent\textbf{Common Failure Patterns. }
We explore the most common failure patterns encountered during large model-based code translation, focusing on their underlying causes, how they manifest in practice, and the strategies needed to address them. By analyzing these recurring issues, we aim to provide actionable insights for improving the accuracy and reliability of cross-language code conversion processes.

\emph{Static Variable Misalignment. }
A frequent issue encountered during translation is the inconsistent handling of static variable naming conventions. For example:

\begin{lstlisting}[language=Java, basicstyle=\footnotesize,frame=single]
public void Stop()
{
    status = GameStatus.Stopped;
}
\end{lstlisting}

The corresponding C\# code raised an error (\texttt{CS0117}) because the enum member \texttt{Stopped} was incorrectly translated. In C\#, enum members are often defined using uppercase conventions, such as \texttt{STOPPED}. This discrepancy arises because Java typically uses mixed-case identifiers, leading to capitalization errors during translation. To avoid such issues, translators should implement specific mappings for capitalization-sensitive identifiers between languages.

\emph{Namespace and Duplicate Definitions. }
Another common error (\texttt{CS0101}) occurs when namespaces contain duplicate definitions due to repetitive code generation. Consider the following Java snippet:

\begin{lstlisting}[language=Java, basicstyle=\footnotesize,frame=single]
public class Candy
{
    public Candy(string flavor) { }
}
\end{lstlisting}

If the translator generates multiple constructors with identical signatures for this class in C\#, the compiler will flag a conflict, as C\# enforces unique member definitions within a namespace or class. The solution involves ensuring that constructors or methods with overlapping signatures are merged or disambiguated during translation.

\emph{Unresolved Names and Contextual Misinterpretations. }
Translation errors often stem from missing imports or incorrect mappings of contextual elements, leading to errors like \texttt{CS0103} (``The name does not exist in the current context''). For example:

\begin{lstlisting}[language=Java, basicstyle=\footnotesize,breaklines=true,frame=single]
private int RandomInt(int min, int max)
{
    return ThreadLocalRandom.Current.Next(min, max + 1);
}
\end{lstlisting}

In this case, the C\# compiler failed because \texttt{ThreadLocalRandom} is not recognized in C\#. Instead, C\# provides a \texttt{Random} class with similar functionality. Translators must correctly identify equivalent libraries and methods in the target language or include necessary imports automatically.

\emph{Undefined Methods. }
Errors such as \texttt{CS1061} occur when the translated code references methods or properties that are undefined in the target language. For instance:

\begin{lstlisting}[language=Java, basicstyle=\footnotesize,frame=single]
_wizards[wizard].SetWisdom(amount);
\end{lstlisting}

This snippet assumes the existence of a method \texttt{SetWisdom} in the \texttt{Wizard} class, but the translator failed to verify its presence. Such semantic gaps can be resolved by enhancing the translator's ability to cross-reference method definitions during conversion and generating warnings for missing methods.

\emph{Runtime Logical Failures. }
Even after resolving compilation errors, logical inconsistencies in the translated code can lead to runtime issues. For example:

\begin{lstlisting}[language=Java, basicstyle=\footnotesize,breaklines=true,frame=single]
private void Register(Weapon weapon, string operation)
{
    if (!_context.TryGetValue(operation, out var weaponsToOperate))
    {
        weaponsToOperate = new List<Weapon>();
    }
    weaponsToOperate.Add(weapon);
    _context[operation] = weaponsToOperate;
}
\end{lstlisting}

Here, a null reference error occurs because the \texttt{\_context} dictionary was not properly initialized before use. Such runtime errors are challenging to detect during static analysis and highlight the need for robust runtime testing frameworks to identify and address logical flaws in translated code.
\subsection{\bench~Effectiveness}
To demonstrate the effectiveness of our method, we conducted two comparative experiments on GPT-4o. These experiments aim to validate (1) the fineness and comprehensiveness of our evaluation mechanism and (2) the necessity of using skeletons during the translation process.
\begin{figure}[t]
  \centering
  \includegraphics[width=0.48\textwidth]{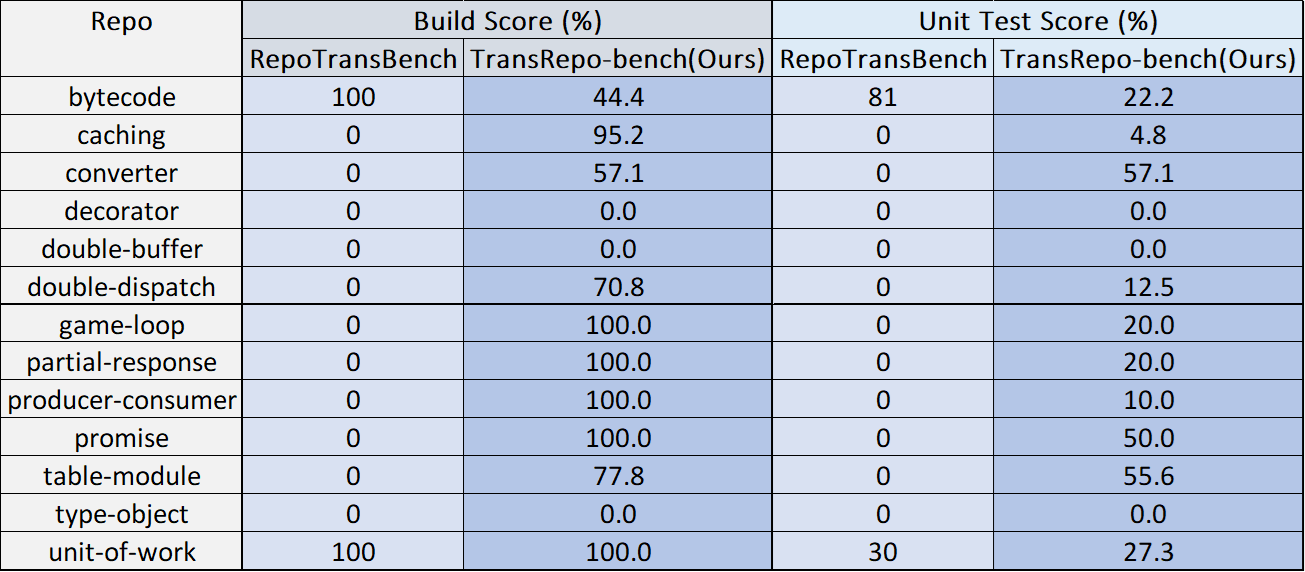}
  \caption{Comparison of RepoTransBench and FineEval evaluation methods on each repository.}
  \label{fig: repotransbench}
\end{figure}
\begin{figure}[t]
  \centering
  \includegraphics[width=0.47\textwidth]{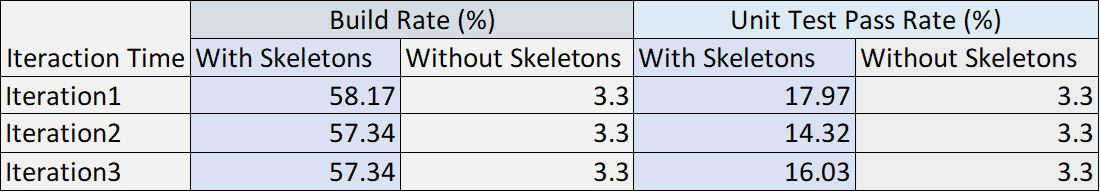}
  \caption{Experiments without Skeletons}
  \label{fig: skeleton-effectiveness}
\end{figure}
\begin{figure}[t]
  \centering
  \includegraphics[width=0.49\textwidth]{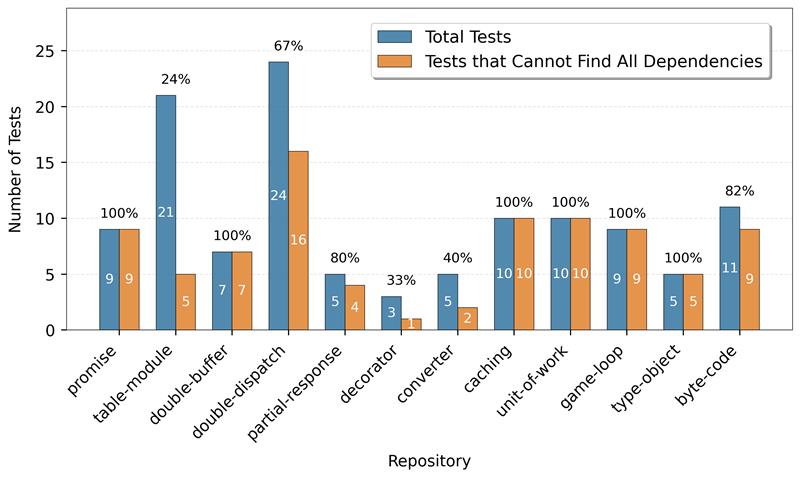}
  \caption{Experiments without Skeletons}
  \label{fig: no-skeletons}
\end{figure}

\subsubsection{Validating the Fineness of Our Evaluation Mechanism}

The first experiment demonstrates that our evaluation mechanism provides a more fine-grained and comprehensive evaluation of a library’s translation quality. Unlike RepoTransBench~\cite{wang24}, which evaluates the entire translated project by directly building and testing it without skeletons, our method scores individual components. This avoids the issue of a single error overshadowing other correct translations.

As shown in Figure \ref{fig: repotransbench}, RepoTransBench~\cite{wang24} achieves a score of 0 for most tasks, with only two out of thirteen tasks being successfully evaluated. In contrast, our method can assign scores to each segment of a project, even in cases where overall compilation fails. For instance, our approach successfully compiles and scores all unit test-related code segments, achieving a 100\% success rate for these cases. This highlights the advantage of our fine-grained evaluation, which ensures partial successes are recognized rather than completely dismissing the translation due to isolated failures.

\subsubsection{Proving the Necessity of Skeletons in Translation}

The second experiment aims to validate the necessity of providing skeletons of the target repository during the translation process. As shown in Figure~\ref{fig: skeleton-effectiveness}, omitting these skeletons substantially degrades both the build success rate and the unit test pass rate across three iterations of our translation pipeline.

The reason for the degradation of the score is caused by the lack of the target repository skeleton in the translation process for the inter-file dependencies and interfaces set in advance, resulting in the failure to find the function under test during the test. As illustrated in Figure~\ref{fig: no-skeletons}, omitting the skeletons results in many dependencies being unresolved during the translation process. This failure renders our scoring mechanism ineffective because unresolved dependencies cause all build and test scores to drop to zero.

For example, the absence of skeletons leads to dependencies being completely unresolvable for certain libraries, as indicated by the high proportion of tests that cannot locate dependencies in the bar chart. This demonstrates that skeletons are essential in the translation process, ensuring that dependencies are correctly identified and enabling the successful evaluation of the translated code.

\noindent\textbf{Summary.}
These experiments collectively establish that our method is superior in two key aspects:
\begin{itemize}[noitemsep, topsep=0pt,leftmargin=*]
\item Our evaluation mechanism is more granular and comprehensive, capturing the quality of translation even when partial failures occur.

\item Providing skeletons during translation is crucial to ensure dependency resolution and enable accurate evaluation.
\end{itemize}

\section{Related Work}

\subsection{Code Translation}

Code translation (source-to-source translation) converts code from one language to another while preserving semantics. Traditional rule-based compilers and intermediate representations (e.g., Babel, Roslyn) work well for constrained cases but falter with complex constructs. Recent AI-driven approaches use neural networks, including sequence-to-sequence models \cite{luong16}, transformers \cite{vaswani23}, and pre-trained models like CodeBERT \cite{feng20}, CodeT5 \cite{wang21}, to improve translation by capturing structural nuances.

Bhattarai et al. \cite{bhattarai24} introduced a few-shot, retrieval-based technique for guiding LLMs in code translation, while Tao et al. \cite{tao24} utilized an intermediary language (Go) to facilitate translations. Unsupervised cross-lingual code representations have also emerged, exemplified by TransCoder \cite{baptiste20}, which handles translations without parallel datasets.

AlphaTrans \cite{ibrahimzada24} is a neuro-symbolic framework for repository-level code translation, employing program analysis and dynamic testing for validation. Shiraishi et al. \cite{shiraishi24} proposed a context-aware C-to-Rust translator, enhancing large-scale compilation success via code segmentation and context-supplementing prompts. Oxidizer \cite{zhang24} likewise addresses repository-level translation with feature mapping, type-compatibility checks, and a semantics-driven phase using unit tests to preserve functionality.

Despite these contributions, AlphaTrans has notable scalability gaps: (1) it handles build errors poorly and lacks comprehensive testing; (2) it overlooks semantic alignment in test case translation, risking untranslated assertions or artificially aligned classes; (3) its fixed-rule approach struggles with special syntax (e.g., method annotations in Java). Our method mitigates these issues by focusing on build error ratios, test case alignment, and more flexible handling of complex syntax.

Many studies \cite{tang23, baptiste20, roziere22, yin24, yang24, jiao23, jana24, di24, tipirneni24, yan23} concentrate on short code from competitive programming \cite{puri21, lu21}, educational websites \cite{yan23, ahmad23}, or custom tasks \cite{liu23, chen21}. Some \cite{Pan24, eniser24, zhang23} address translating longer code (100+ lines) but report limited success. Novel training strategies \cite{baptiste20, roziere22, szafraniec23, jana24, tipirneni24} could boost the LLM in our approach, while prompting \cite{tang23} and repair methods \cite{yin24} are also relevant. Automated program repair techniques \cite{xia23, kong24} may further address I/O equivalence issues if adapted to translation-specific errors. SYZYGY \cite{shetty24} translates C to safe Rust by combining LLM-driven code/test generation with dynamic analysis to ensure correctness.

\subsection{Code Translation Benchmarks}

Benchmarks are crucial for evaluating the performance of code translation systems. Early benchmarks consisted of manually curated small-scale function pairs, which were limited in scale and diversity. Modern benchmarks have expanded to include large-scale datasets with a variety of programming languages, encompassing both open-source projects and synthetic code samples.

AdvBench \cite{robey21} is a benchmark designed for evaluating TransCoder, including programs written in Java, C++, and Python. It assesses translation fidelity using metrics such as BLEU, Exact Match (EM), and Code Execution Accuracy, focusing on real-world applicability. Similarly, CodeNet \cite{puri21} is a vast dataset containing 14 million code samples across 50 programming languages, providing a comprehensive foundation for training and evaluating code translation models.

Beyond general-purpose benchmarks, task-specific benchmarks address specialized domain challenges. For instance, the CodeXGLUE benchmark \cite{lu21} evaluates various programming tasks, including code translation, by incorporating execution-based metrics to ensure functional correctness. However, these benchmarks often lack coverage for niche languages, complex system-level code, or real-world constraints like incomplete libraries and ambiguous syntax.

RustRepoTrans \cite{ou24} is the first benchmark to include repository-level dependencies for Rust code translation, addressing the limitations of function-level datasets. Evaluations using RustRepoTrans revealed a significant performance drop (41.5\%-56.2\%) when handling repository-level tasks, highlighting the difficulties in managing dependencies and cross-file interactions in real-world scenarios. Similarly, RepoTransBench \cite{wang24} is a benchmark for repository-level code translation that features 100 repositories with automated test suites to evaluate translation quality. It tackles challenges such as complex configurations, resource file handling, and test case migration.

However, RepoTransBench has certain limitations that our approach overcomes:
(1) Lack of Skeleton Framework: RepoTransBench does not utilize skeletons, making it difficult to constrain the interfaces generated by large models. This often leads to interface misalignments during testing and restricts their use in incremental translation scenarios. In contrast, our skeleton-based approach ensures tighter control and better adaptability.
(2) Absence of Test Checking: RepoTransBench lacks a robust mechanism for verifying test results. Our method ensures alignment by running unit tests on both source and target language skeletons, providing a more reliable evaluation process.
(3) Coarse-Grained Evaluation: RepoTransBench executes unit tests directly without isolating dependencies, which can result in compounded translation errors affecting test outcomes. Our approach isolates relevant dependencies within the skeleton, allowing for finer-grained evaluation of translation quality and minimizing the impact of such errors.

\section{Conclusions}
This paper presents \bench, a novel benchmark and framework addressing critical challenges in repository-level code translation, including inter-module coherence, dependency management, and fine-grained evaluation. By leveraging a two-step translation approach centered on repository skeletons, we ensure structural consistency while enabling precise and incremental translation. The proposed fine-grained evaluation mechanism, which scores translation quality at the unit test level, offers detailed feedback beyond traditional binary metrics.

We validate the effectiveness of skeleton-based translation and fine-grained evaluation, demonstrating that incorporating repository skeletons significantly improves translation accuracy by resolving inter-file dependencies and enabling partial validation. Comprehensive experiments on state-of-the-art large language models reveal key challenges in repository-level translation, such as error propagation and runtime failures. Our failure analysis highlights common pitfalls, including static variable misalignment, unresolved dependencies, and namespace conflicts, offering actionable insights for enhancing cross-language translation systems.

\bibliography{ref}


\end{document}